\def\thetac{\theta_C}

\def\dnuup{\partial^{\nu}}

\def\dsone{|\Delta S|=1}

\def\smalllinestretch{\renewcommand{\baselinestretch}{0.7}}
\def\largelinestretch{\renewcommand{\baselinestretch}{1.0}}
 \def\tr {\, \mbox{tr} \,}

 \def\im {\, \mbox{Im} \,}
 \def\re {\, \mbox{Re} \,}
 \def\hc {   \mbox{h.c.} }
 \def\piz{\pi^0}

 \def\et{\eta}
 \def\ep{\eta'}

 \def\mix{(\( \piz\!, \et, \ep\))-mixing}
 \def\A{{\cal A}}
 \def\B{{\cal B}}
 \def\L{{\cal L}}
 \def\M{{\cal M}}
 
 \def\O{{\cal O}}

 \def\qq{<\!\!\bar{q} q\!\!>}

 \def\fz{F_0}

 \def\up{U^+}
 \def\fj{\varphi}
 \def\Fi{\Phi}

 \def\dmu{\partial_{\mu}}
 \def\dnu{\partial_{\nu}}

 \def\dii{\partial^2}

 \def\dmuup{\partial^{\mu}}
 \def\dnuup{\partial^{\nu}}

 \def\epsp {|{\varepsilon^{'}\! /   \varepsilon}|}
 \def\epspp{|{\varepsilon^{'} \over \varepsilon}|}

 \def\diag{\mbox{diag}}
 \def\mev{\,\mbox{MeV}}
 \def\gev{\,\mbox{GeV}}

 \def\kpmtoppp{\(K^\pm \to 3\pi\)}
 \def\delg{\Delta g(K^\pm\to 3\pi)}
 \def\gt{\widetilde{G}}

\newcommand{\su}[1]{\mbox{\small\it #1}}
\documentstyle[11pt,a4]{article}

\def\largelinestretch{\renewcommand{\baselinestretch}{1.5}}
\largelinestretch\small\normalsize

\def\largelinestretch{\renewcommand{\baselinestretch}{1.2}}
\title{
                On the Origin of the Enhancement\\
        of CP-violating Charge Asymmetries in \kpmtoppp{} Decays\\
               Predicted from Chiral Theory
      }
 \author{
A.A.Bel'kov${}^1$,
G.Bohm${}^2$,
D.Ebert${}^3$,
A.V.Lanyov${}^1$,
A.Schaale${}^2$
\\
\\
\small
${}^1$
        Particle Physics Laboratory, Joint Institute for Nuclear Research,
\hfill\\
\small
        Head Post Office, P.O.Box 79, 101000 Moscow, Russia
\hfill\\
\small
${}^2$
        DESY-Institute for High Energy Physics,
        Platanenallee 6, O-1615 Zeuthen, Germany
\hfill\\
\small
${}^3$
        Institut f\"ur Elementarteilchenphysik, Humboldt-Universit\"at,
\hfill\\
\small
        Invalidenstra\ss e 110, O-1040 Berlin, Germany
\hfill\\
}
\begin{document}
\largelinestretch\normalsize
   \thispagestyle{empty}
   \begin{titlepage}
   \maketitle
   \begin{abstract}
   We present an analysis of the enhancement of CP-violating
charge asymmetries in \kpmtoppp{} decays.
   Calculations of decay amplitudes are performed on the basis of
bosonized strong and weak Lagrangians derived from QCD-motivated
quark Lagrangians.
   We show that the interplay of fourth-order contributions of
chiral Lagrangians for strong interactions and penguin operators
in weak interactions significantly enhances the charge asymmetries.
\vfill
   Contributed paper to XXVI International Conference on High Energy
Physics, Dallas, Texas, August 6, 1992.

   \end{abstract}
   \end{titlepage}
   %
   %
   %

   Recently, much interest is devoted to the question of a possible
enhancement of direct CP-violation effects in \kpmtoppp{} decays first
proposed in \cite{pl2}. In this paper from chiral Lagrangians with
fourth-order derivative terms, including meson loop rescattering
effects, the CP-violating charge asymmetry of the Dalitz plot slope
parameter, $\delg$, was estimated to be of the order of $10^{-3}$.
   On the other hand, in several recent papers
\cite{cheng,dambrosio,isidori,shabalin} estimates have been given
which are 1-2 orders of magnitude smaller.
   By this reason and due to the fact that this problem is surely of
great importance for the choice of the future experimental program at
$\phi$- and $K$-factories we have reanalyzed this question
within our approach taking into account additional effects.
   First we will give the main definitions and assumptions used,
next display all parameters entering the calculation and then show
step by step the effects of including various refinements:
fourth-order derivative terms, additional fourth-order $s$-quark mass
terms, \mix{}, one-loop corrections connected to meson rescattering
and electromagnetic penguins.
   In this way we hope to demonstrate more completely the mechanisms of
enhancement for the effects of direct CP-violation in \kpmtoppp{} decays.
For simplicity, we will consider only the above mentioned charge
assymetry of the slope parameter
\[
 \delg = {g(K^+)-g(K^-) \over g(K^+)+g(K^-)}.
\]
   The enhancement effects for other asymmetries, e.g. of branching
ratios, are of the same origin.

   The starting point of the following estimates is the
effective Lagrangian describing nonleptonic weak interactions
with strangeness change $\dsone{}$ which is given on the quark
level by \cite{vzsh,gilman-wise,bijnens-wise}:
 \begin{equation}
 \L^{\su{nl}}_{\su{w}} = \gt \sum_{i=1}^8 c_i \, {\cal O}_i \; .
 \label{weak-lagr}
 \end{equation}
   Here $ \gt = \sqrt2 \, G_F \, \sin \thetac \,\cos \thetac$ is the
weak coupling constant;
   $c_i$ are Wilson coefficient functions which may be calculated in
the QCD leading-log approximation, depending then explicitly on the
renormalization scale $\mu$.
${\cal O}_i$ are the four-quark operators
consisting of products of left- and/or right-handed quark currents:
 \begin{eqnarray}
\O_1 &=&   \bar{u}_L \gamma_\mu u_L \; \bar{d}_L \gamma^\mu s_L
       -   \bar{d}_L \gamma_\mu u_L \; \bar{u}_L \gamma^\mu s_L,
\nonumber\\
\O_2 &=&   \bar{u}_L \gamma_\mu u_L \; \bar{d}_L \gamma^\mu s_L
       +   \bar{d}_L \gamma_\mu u_L \; \bar{u}_L \gamma^\mu s_L
       + 2 \bar{d}_L \gamma_\mu d_L \; \bar{d}_L \gamma^\mu s_L
       + 2 \bar{s}_L \gamma_\mu s_L \; \bar{d}_L \gamma^\mu s_L,
\nonumber\\
\O_3 &=&   \bar{u}_L \gamma_\mu u_L \; \bar{d}_L \gamma^\mu s_L
       +   \bar{d}_L \gamma_\mu u_L \; \bar{u}_L \gamma^\mu s_L
       + 2 \bar{d}_L \gamma_\mu d_L \; \bar{d}_L \gamma^\mu s_L
       - 3 \bar{s}_L \gamma_\mu s_L \; \bar{d}_L \gamma^\mu s_L,
\nonumber\\
\O_4 &=&   \bar{u}_L \gamma_\mu u_L \; \bar{d}_L \gamma^\mu s_L
       +   \bar{d}_L \gamma_\mu u_L \; \bar{u}_L \gamma^\mu s_L
       -   \bar{d}_L \gamma_\mu d_L \; \bar{d}_L \gamma^\mu s_L,
\nonumber\\
\O_5 &=& \bar{d}_L \gamma_\mu \lambda^a_c s_L
  \left( \sum_{q=u,d,s} \bar{q}_R \, \gamma^\mu \, \lambda^a_c \, q_R \right),
\quad
\O_6 = \bar{d}_L \gamma_\mu s_L
  \left( \sum_{q=u,d,s} \bar{q}_R \, \gamma^\mu \, q_R \right),
\nonumber\\
\O_7 &=& \bar{d}_L \gamma_\mu s_L
  \left( \sum_{q=u,d,s} \bar{q}_R \, \gamma^\mu \, Q \, q_R \right),
\quad
\O_8 = \bar{d}_L \gamma_\mu \lambda^a_c s_L
  \left( \sum_{q=u,d,s} \bar{q}_R \, \gamma^\mu \, \lambda^a_c \, Q \, q_R
\right),
\nonumber
 \end{eqnarray}
where $q_{L,R} = \frac12 \, (1\mp\gamma_5) q$; $\lambda^a_c$ are
the generators of the $SU(N_c)$ color group; $Q$ is the matrix of
electric quark charges.
   The operators ${\cal O}_{5,6}$ containing right-handed
currents are generated by gluonic penguin diagrams and the analogous
operators ${\cal O}_{7,8}$ arise from electromagnetic
penguin diagrams.

   The bosonized version of the effective Lagrangian (\ref{weak-lagr}) can be
expressed in the form \cite{91-08}:
 \def\jim{(J^1_{L\mu} - i J^2_{L\mu})}
 \def\rim{(J^1_R - i J^2_R)}
 \def\jiii{(J^3_{L\mu} + \frac1{\sqrt3} \, J^8_{L\mu})}
 \def\riii{(J^3_R - \frac1{\sqrt3} \, J^8_R - \sqrt\frac23 \, J^0_R)}
 \def\jivp{(J^4_{L\mu} + i J^5_{L\mu})}
 \def\livp{(J^4_L + i J^5_L)}
 \def\jvip{(J^6_{L\mu} + i J^7_{L\mu})}
 \def\lvip{(J^6_L + i J^7_L)}
 \def\rvip{(J^6_R + i J^7_R)}
 \begin{eqnarray}
 {\cal L}^{\su{nl}}_{\su{eff}} &=& \gt \Bigg\{
  (-\xi_1 + \xi_2 + \xi_3) \bigg[ \jim \jivp - \jiii \jvip \bigg]
\nonumber\\
&&+ (\xi_1 + 5\,\xi_2) \sqrt\frac23 \, J^0_{L\mu} \jvip
+ {10 \over \sqrt3} \, \xi_3 \, J^8_{L\mu} \jvip
\nonumber\\
&&+ \xi_4 \bigg[ \jim \jivp + 2 \, J^3_{L\mu} \jvip \bigg]
\nonumber\\
&&- 4 \, \xi_5 \bigg[ \rim \livp - \riii \lvip
\nonumber\\
&&\qquad - \sqrt\frac23 \, \rvip (\sqrt2 J^8_L - J^0_L)
           \bigg]
\nonumber\\
&&+ \xi_6 \, \sqrt\frac32 \, \jivp J^0_R
+ 6 \, \xi_7 \, \jvip (J^3_{R\mu} + \frac1{\sqrt3} \, J^8_{R\mu})
\nonumber\\
&&- 16 \, \xi_8 \, \bigg[ \rim \livp + \frac12 \, \riii \lvip
\nonumber\\
&&\qquad + \frac1{\sqrt6} \, \rvip (\sqrt2 \, J^8_L - J^0_L)
            \bigg]
 \Bigg\} + \hc \, .
\nonumber
 \end{eqnarray}
   Here $J^a_{L/R \, \mu}$ and $J^a_{L/R}$ are bosonized $(V\mp A)$
and $(S\mp P)$ meson currents corresponding to the quark
currents
$\bar q \gamma_\mu \frac14 (1\mp \gamma^5) \lambda^a q$
and
$\bar q \frac14 (1\mp \gamma^5) \lambda^a q$,
respectively ($\lambda^a$ are the generators of the $U(3)_F$ flavor group);
\begin{eqnarray}
&&\xi_1 = c_1 \left( 1 - {1 \over N_c} \right), \qquad
\xi_{2,3,4} = c_{2,3,4} \left( 1 + {1 \over N_c} \right),
\nonumber\\
&&\xi_{5,8} = c_{5,8} + {1 \over 2 N_c} c_{6,7}, \qquad
\xi_{6,7} = c_{6,7} - {2 \over N_c} c_{5,8},
\end{eqnarray}
where the color factor ${1/N_c}$ originates from the Fierz-transformed
contribution to the nonleptonic weak effective chiral Lagrangian \cite{91-08}.

   The meson currents can be derived from the $p^2$- and $p^4$-terms
of the quark determinant in a QCD-motivated chiral quark model using
the bosonization procedure described in \cite{91-08}:
\begin{eqnarray}
J_{L\,\mu}^{a\,(p^2)} &=&
i\, \frac{F^2_0}{8} \tr \left[ \lambda^a \dmu U \,\up(1 + {1 \over 2m^2} \M)
                                                                    \right] +
\hc,
\nonumber \\
{J}^{a\,(p^2)}_{L} &=&
\frac{F^2_0}{4} \, m \, R \, \tr (\, \lambda^a \up)
+ \frac{F^2_0}{8m} \, \tr \left[ \lambda^a (\dii \up + 2 \, \up \M) \right],
\nonumber \\
{J}^{a\,(p^4)}_{L\, \mu} &=&
{i \, N_c \over 64\pi^2} \, \tr \Bigg\{ \lambda^a
 \bigg[ \frac13 \, \dnu U \, \dmu \up \dnuup U \up
+ \{\M, \, \dmu U\} \, \up
\nonumber \\
&&- \frac1{6m^2} \Big(
            \{\dmu \dii U, \, \M \} \, \up
           - [\dmu \dnu U \, \dnuup \up, \, \M ]
           + \dnu U \{\dmu \dnu \up, \, \M \}
                \Big)
                \bigg]
                \Bigg\} + \hc ,
\nonumber \\
{J}^{a\,(p^4)}_{L} &=&
-{N_c \over 192\pi^2m} \, \tr \Bigg\{ \lambda^a
 \bigg[ 3 \Big(
                -m^2 (\up\, \dmu U\, \dmuup \up + \dmu\up \, \dmuup U \, \up)
                + \{ \dii \up, \, \M\}
          \Big)
\nonumber \\
&&+ \frac12 \Big(
           \up \, ( \{\M, \, \dmu \up \dmuup U\} + \dmu \up \, \M \, \dmuup U)
\nonumber \\ && \qquad \quad
          +( \{\dmu U \dmuup \up, \, \M \} + \dmu U \, \M \, \dmuup \up) \up
         \Big)
         \bigg]
         \Bigg\}.
\label{weak-cur}
\end{eqnarray}
Here $U = \Omega^2$,
$\Omega = \exp \left\{ {i \Phi \over \sqrt2 \, F_0} \right\}$,
where $\Fi = \frac1{\sqrt3}\fj_0 + \frac1{\sqrt2} \sum_{a=1}^8 \lambda_a\fj_a$
is the matrix of pseudo- scalar meson fields $\fj_a$.
   $F_0$ is the (bare) decay constant of the $\pi \to \mu\nu$ decay.
   The terms containing the matrix $\M = 2m\, \Omega \, m_0\,\Omega^+$,
$m$ being the average constituent quark mass and
$m_0 = \diag(m^0_u, m^0_d, m^0_s)$ is the mass matrix of current quarks,
take into account the additional contributions from the quark mass
expansion which are dominated by $s$-quark mass terms.
   The contributions of the gluon and electromagnetic penguin
operators are determined by the parame- ter
\[
R =
     \ {\qq \over mF_0^2}
\]
 where $\qq$ is the quark condensate.

   The corresponding contributions to the
$p^2$- and $p^4$-parts of the effective Lagrangian of strong
interaction are \cite{91-08,ebert-reinhardt}
\footnote {The notation $p^2(p^4)$-terms refers to $p^2(p^4)$-order in
the corresponding momentum expansion of the quark determinant
including terms with additional mass dimensions from $\M$. }
\begin{eqnarray}
 \L^{(p^2)} &=& {F_0^2 \over 4} \, \tr(\dmu U \, \dmuup \up)
 + {F_0^2 \over 4} \, \tr
      \left[ M
               \left(1 + {F_0^2 \over 2m \qq} \, \dii
               \right) U + \hc
      \right],
\nonumber \\
 \L^{(p^4)} &=& {N_c \over 32\pi^2} \, \tr
      \left[ \frac16 \, \big(\dmu U \, \dnu \up \big)^2
            - {m \, F_0^2 \over \qq}
      \left(1 - {4 \pi^2 \, F_0^2 \over m^2 N_c} \right)
                              \dmu U \dmuup \up \big(M\, U + \up \, M\big)
      \right]
\label{strong-lagr}
\end{eqnarray}
where $M = \diag(\chi_u^2, \chi_d^2, \chi_s^2)$,
$\chi_i^2 
 = -2 m_i^0 \, F_0^{-2} \qq$.

     The parameters $\chi_i^2$, $m_i^0$ and $m$ have been fixed in
\cite{91-08} by the spectrum of pseudoscalar and vector mesons.
Here we use the value $m=380 \, \mev$
and the relation
$m_s^0 = \widehat{m}_0 \,\, \chi_s^2 / m_\pi^2$
where
$\widehat{m}_0 \equiv (m_u^0+m_d^0)/2 \approx 5 \, \mev$ and
$\chi_u^2 = 0.0114\,\gev^2, \quad \chi_d^2 = 0.025\,\gev^2, \quad
   \chi_s^2 = 0.47\,\gev^2.$
   Taking into account the additional Goldberger-Treiman
contribution to $F_{K,\pi}$ arising from current quark mass
splitting the value $F_0=89\,\mev$ was obtained in \cite{91-08}.
\footnote { Note the appearance of the important new symmetry breaking
term $\sim \,tr M \dii U $ in (\ref{strong-lagr}) which contributes to
$F_K - F_{\pi}$ splitting. This term plays the major role for the
effect of direct CP-violation considered here. }

   The experimental status of the first $p^4$-order term of the strong
Lagrangians (\ref{strong-lagr}) was discussed in \cite{strong,our}
on the basis of the analysis of data on $d$-wave $\pi\pi$-scattering
and the decay width of $\eta^{'} \to \eta\,2\pi$.
   In the present work we again will drop tachyonic contributions to both
the strong Lagrangian $\L^{(p^4)}$ and the bosonized currents
$J_{L\,\mu}^{a\,(p^4)}$ and
$J_L^{a\,(p^4)}$.

   Using isospin relations, the $K \to 2\pi$ and $K\to 3\pi$ decay
amplitudes can be parametrized as
\begin{eqnarray}
T_{K^+ \to \pi^+ \pi^0} &=& {\sqrt3 \over 2}\, A_2,
\nonumber \\
T_{K^0_S \to \pi^+ \pi^-} &=& \sqrt{2 \over 3}\, A_0 + {1 \over \sqrt3}\, A_2,
\qquad
T_{K^0_S \to \pi^0 \pi^0} = \sqrt{2 \over 3}\, A_0 - {2 \over \sqrt3}\, A_2,
\end{eqnarray}
and
\begin{eqnarray}
T_{K^+ \to \pi^+ \pi^+ \pi^-} &=& 2 \, (\A_{11}+\A_{13})
                                 -Y \, (\B_{11}+\B_{13}-\B_{23}) + O(Y^2),
\nonumber \\
T_{K^+ \to \pi^0 \pi^0 \pi^+} &=& (\A_{11}+\A_{13})
                                 +Y \, (\B_{11}+\B_{13}+\B_{23}) + O(Y^2),
\end{eqnarray}
   where $Y = (s_3-s_0)/m_\pi^2$ is the Dalitz variable and
$s_i={(k-p_i)}^2$, $s_0=m_K^2/3 + m_\pi^2$;
$k$, $p_i$ are four-momenta of the kaon and $i$th pion ($i=3$ belongs
to the odd pion).
   The Dalitz-plot distribution can be written as a power series
expansion of the amplitude squared, $|T|^2$, in terms of the
corresponding kinematical variables $Y$ and $X$
\begin{equation} \label{expanxy}
|T|^2 \propto 1 + gY + hY^2 + kX^2 + ...
\end{equation}
where $X = (s_2-s_1)/m_\pi^2$; $g$, $h$ and $k$ are the slope parameters.

   The isotopic amplitudes $A_{2,0}$ determine the $K \to 2\pi$
transitions into states with isospin $I=2,0$, respectively:
\[ A_2 = a_2 \, e^{i\delta_2}, \qquad
   A_0 = a_0 \, e^{i\delta_0}
\]
where $\delta_{2,0}$ are the phases of $\pi\pi$-scattering.
   It is well known, that direct CP-violation results in an additional
(small) relative phase between $a_2$ and $a_0$.

   The isotopic amplitudes $\A_{IJ}$, $\B_{IJ}$ of $K\to3\pi$
decays have two indices: $I$, the isospin of the final state, and
$J$, the doubled value of isospin change between the initial and
final states.
   It is customary also for the $3\pi$-system to introduce strong
phase shifts $\alpha_1$, $\beta_1$ and $\beta_2$ corresponding to the
relevant isospin states $I=1_s$ (symmetric), $I=1_m$ (mixed symmetric),
$I=2$ by writing \cite{devlin}
\[ \A_{11} + \A_{13} = (a_{11} + a_{13}) \, e^{i\alpha_1}, \qquad
   \B_{11} + \B_{13} = (b_{11} + b_{13}) \, e^{i\beta_1}, \qquad
   \B_{23} = b_{23} \, e^{i\beta_2}.
\]
   We shall use this representation here only in order to display more
cleary the relationships between the main contributions to the direct
CP-violation effect.
   Because the strong Hamiltonian is not necessarily diagonal with
respect to the $I=1_s, I=1_m$ isospin states and, if isospin breaking
is included, even $I=1$ and $I=2$ states get mixed, leading to the
necessity of introducing more phases (cf \cite{pl2}), the exact
calculations of $\delg$ have to be done using the
complex quantities $\A_{IJ}$, $\B_{IJ}$ given below by (\ref{defab})
directly, without introducing the strong phases $\alpha_1$,
$\beta_{1,\, 2}$ explicitly.

   Let us next introduce the contributions of the four-quark
operators $\O_i$ to the isotopic amplitudes
$A_I^{(i)}$, $\A_{IJ}^{(i)}$ and $\B_{IJ}^{(i)}$ by the relations
\begin{eqnarray}
A_2&=&-i\,\sum_{i=1}^8
\xi_i\,{\sqrt3\over2}\,\gt\,\fz\,(m_K^2-m_\pi^2)\,A_2^{(i)},
\qquad
A_0=-i\,\sum_{i=1}^8
\xi_i\,\sqrt{\frac38}\,\gt\,\fz\,(m_K^2-m_\pi^2)\,A_0^{(i)};
\nonumber \\
\A_{IJ}&=&-\sum_{i=1}^8
\xi_i\,\left(\gt\,{m_K^2-m_\pi^2\over12}\right)\,\A_{IJ}^{(i)},
\qquad
\B_{IJ}=-\sum_{i=1}^8 \xi_i\,\left(\gt\,{m_\pi^2\over4}\right)\,\B_{IJ}^{(i)}.
\label{defab}
\end{eqnarray}

   The parameter $\varepsilon^{'}$ of direct CP-violation in
$K\to2\pi$ decays and the charge asymmetry of the slope parameters
$\delg$ can be expressed by the formulae
\begin{equation}
\varepsilon^{'} =  - {\omega\over\sqrt2}\,{\im a_0\over\re a_0}
                  \left(1 - {1\over\omega}\,{\im a_2\over\im a_0}\right)\,
                        \exp[i(\pi/2+\delta_2-\delta_0)]
\label{eps}
\end{equation}
and
\begin{equation}
\Delta g \left( K^\pm \to
                          \left\{ {\Big.}^{\mbox{$\pi^\pm \pi^\pm \pi^\mp$}}
                                         _{\mbox{$\pi^0   \pi^0   \pi^\pm$}}
                          \right\}
         \right)
 = { \im F_1\, \sin(\alpha_1-\beta_1) \mp
     \im F_2\, \sin(\alpha_1-\beta_2) \over
     \re F_1\, \cos(\alpha_1-\beta_1) \mp
     \re F_2\, \cos(\alpha_1-\beta_2) }
\label{delg}
\end{equation}
where $\omega=\re a_2 / \re a_0$;
$F_1=(a_{11}^* + a_{13}^*)(b_{11}+b_{13})$,
$F_2=(a_{11}^* + a_{13}^*)b_{23}$.
\footnote {In deriving charge asymmetries, one has to keep in mind,
that charge conjugation does reverse the phases of $\xi_i$ but not those
$A_I^{(i)}$, $\A_{IJ}^{(i)}$, $\B_{IJ}^{(i)}$. }

   Using only the currents $J^{a\,(p^2)}_{L\,\mu}$,
$J^{a\,(p^2)}_L$ and the $p^2$-order part of the strong Lagrangian $\L^{(p^2)}$
it is possible to reproduce the following relations between $K\to2\pi$,
$K\to3\pi$ isotopic amplitudes in the spirit of the ``soft pion''
limit:
\begin{equation}
\A_{11}^{(i)} = \B_{11}^{(i)}=A_0^{(i)}, \qquad
\A_{13}^{(i)} = A_2^{(i)} \qquad (i=1,2, \, \ldots, 6),
\label{soft-pion1}
\end{equation}
where for nonvanishing amplitudes we have
\begin{equation}
A_0^{(1)} = -A_0^{(2,3)} = -1 = -A_2^{(4)}, \qquad
A_0^{(5)} = 4 R .
\label{soft-pion2}
\end{equation}
Moreover,
\begin{eqnarray}
\A_{11}^{(7)} &=& - \frac15 \B_{11}^{(7)} = A_0^{(7)} = -A_2^{(7)} = 2;
\nonumber \\
\A_{11}^{(8)} &=& -4 A_0^{(8)} = 3 \A_{13}^{(8)} = -32\,R(2m^2 R - m_K^2
-m_\pi^2)/(m_K^2-m_\pi^2),
\nonumber \\
A_2^{(8)} &=& 8\,R(m^2 R - m_\pi^2)/(m_K^2-m_\pi^2);
\nonumber \\
\B_{13}^{(4)} &=& -\frac14(5 m_K^2 - 14 m_\pi^2)/(m_K^2-m_\pi^2),
\quad
\B_{13}^{(7)} = -\frac12 (7 m_K^2 + 2  m_\pi^2)/(m_K^2-m_\pi^2),
\nonumber \\
\B_{13}^{(8)} &=&  -2\,R(9m^2 R - 8 m_K^2 -m_\pi^2)/(m_K^2-m_\pi^2);
\nonumber \\
\B_{23}^{(4)} &=&  \frac94 (3 m_K^2 - 2 m_\pi^2)/(m_K^2-m_\pi^2),
\quad
\B_{23}^{(7)} =  \frac92 (m_K^2 - 2 m_\pi^2)/(m_K^2-m_\pi^2),
\nonumber \\
\B_{23}^{(8)} &=&  -18\,R (m^2 R - m_\pi^2)/(m_K^2-m_\pi^2).
\nonumber
\end{eqnarray}
(the other isotopic amplitudes vanish).

   As the first step, let us consider only the contributions due to
the operators $\O_1$, ..., $\O_6$, neglecting the electromagnetic
penguin operators $\O_{7,8}$.
   It is convenient to present the terms in the numerator of the
right-hand side of eq.(\ref{delg}) for $\delg$ in a more visual form
\begin{eqnarray}
\im F_1 &=& \Delta^{(1/2, \, 1/2)} + \Delta^{(1/2,\, 3/2)},
\nonumber \\
\Delta^{(1/2, \, 1/2)} &=& \re a_{11} \, \im b_{11} - \im a_{11} \, \re b_{11},
\qquad
\Delta^{(1/2, \, 3/2)} = \re a_{13} \, \im b_{11} - \im a_{11} \, \re b_{13};
\nonumber \\
\im F_2 &=& -\im a_{11} \, \re b_{23} \equiv \Delta^{'(1/2, \, 3/2)}
\label{delta}
\end{eqnarray}
where $\Delta^{(1/2, \, 1/2)}$ describes the contribution of
the interference of isotopic amplitudes $a_{11}$ and $b_{11}$ for
transitions with $\Delta I=1/2$, and
$\Delta^{(1/2, \, 3/2)}$, $\Delta^{'(1/2, \, 3/2)}$ are the
contributions from interferences of amplitudes $a_{IJ}$ and $b_{IJ}$ with
$\Delta I=1/2$ and $3/2$.
  In writing eq.(\ref{delta}) we assume that direct CP-violation arises
only due to the imaginary parts of the isotopic amplitudes with $\Delta
I=1/2$ generated by the imaginary part of the Wilson coefficient $c_5$
of the penguin operator $\O_5$.
   The contribution of the operator $\O_6$ is small and is therefore neglected.
   From (\ref{soft-pion1}) and (\ref{soft-pion2}) it is obvious that
in the soft-pion limit valid for $p^2$-order terms
$\Delta^{(1/2, \, 1/2)}=0$, and only interferences of
amplitudes with $\Delta I=1/2$ and $3/2$ can contribute to the  charge
asymmetry $\Delta g$ in this limit.

   It is well known that the fourth-order terms of the chiral
Lagrangian $\L^{(p^4)}$ (\ref{strong-lagr}) and their contributions to
the currents $J_{L\,\mu}^{a\,(p^4)}$ and  $J_L^{a\,(p^4)}$ (\ref{weak-cur})
lead to a modification of soft-pion relations
for isotopic $K\to2\pi$ and $K\to3\pi$ amplitudes.
   This modification was already discussed in ref.{} \cite{our}.
   In particular, the additional contributions due to the first term
of  $J_L^{a\,(p^4)}$ to $\A_{11}^{(i)}$ and
$\B_{11}^{(i)}$ ($i$=1, 2, 3, 5)
are:\footnote{The $p^4$-interactions do not contribute to $K\to2\pi$ decays.}
\begin{eqnarray}
\Delta\A_{11}^{(1)}&=&-\Delta\A_{11}^{(2,3)}=-{m_K^2-3m_\pi^2\over12\,F_0^2\pi^2},
\qquad
\Delta\A_{11}^{(5)}= -4R\, {m_K^2-3m_\pi^2 \over 12\,F_0^2\pi^2};
\nonumber \\
\Delta\B_{11}^{(1)}&=&-\Delta\B_{11}^{(2,3)}={m_K^2+3m_\pi^2\over12\,F_0^2\pi^2},
\qquad
\Delta\B_{11}^{(5)}= 4R\, {m_K^2+3m_\pi^2 \over 12\,F_0^2\pi^2}.
\label{p4-correct}
\end{eqnarray}

   One can see from (\ref{soft-pion1}) and (\ref{soft-pion2}) that the
contribution of the Lagrangian $\L^{(p^2)}$ and the corresponding currents
$J_{L\,\mu}^{a\,(p^2)}$ and  $J_L^{a\,(p^2)}$
to the amplitudes with $|\Delta I|=1/2$ is proportional to
$(-\xi_1+\xi_2+\xi_3+4R\,\xi_5)$.
   At the same time the contribution associated to the corrections
(\ref{p4-correct}) is proportional to $(-\xi_1+\xi_2+\xi_3-4R\,\xi_5)$.
   So, $\xi_5$ cannot be absorbed by a redefinition of the parameters $\xi_i$.
   Due to this reason it is possible to separate penguin
and nonpenguin contributions in a data fit for $K\to2\pi$ and
$K\to3\pi$ decays.
   This different behaviour between penguin and
nonpenguin contributions, arising on the fourth-order level, leads to a
nonzero value of the $\Delta^{(1/2,1/2)}$ term which, as will be shown
below, becomes the main source of the enhancement of the CP-violating charge
asymmetry of the slope parameters $\delg$ discussed in this paper.

   Besides $p^4$-interactions, \mix{} and
one-loop corrections corresponding to meson rescattering also
modify the soft-pion relations for the isotopic $K\to2\pi$ and
$K\to3\pi$ amplitudes.
   The results of lengthy symbolic calculations using the
package of tools \cite{reduce} based on the REDUCE system are shown
in table 1.
   Even though the contributions of \mix{} are proportional to the
small mass difference of $d$- and $u$-quarks (breaking of
isotopic symmetry), they give a contribution of about $30\%$ in
the description of direct CP-violation in $K\to2\pi$, $K\to3\pi$
decays due to the new additional contributions to $\im a_2$,
$\im a_{13}$ and $\im a_{23}$ (via $|\Delta I|=3/2$ transitions)
arising from the penguin operator $\O_5$.
   The importance of one-loop corrections for a correct estimation
of direct CP-violation is determined by both the phase shifts
$\alpha_1$, $\beta_1$ and $\beta_2$, induced by
$\pi\pi$-interactions  in final states, and the modification of the real
parts of $A_I$, $\A_{IJ}$ and $\B_{IJ}$ due to $\pi\pi$-, $\pi
K$- and $\bar{K}K$-scattering.

   In our approach the meson loops were estimated using a
special superpropagator regularization method \cite{volkov}
which is particularly well-suited for treating loops in
nonlinear chiral theories.
   The result is equivalent to dimensional regularization, the
difference being that the scale parameter $\mu$ is no longer
free but fixed by the inherent scale of the chiral theory, namely
$\mu = 4\pi \, F_0$, and UV divergences have to be replaced by a finite term
through the substitution
\begin{equation} \label{spprescr}
(C - 1/\varepsilon) \quad \longrightarrow \quad C_{SP}
=2C+1+\frac12\,\left[{d \over dz}
                     \left(\log \Gamma^{-2}(2z+2)
                     \right)
               \right]_{z=0}=-1+4C
\approx 1.309,
\end{equation}
   where $C=0.577$ is the Euler constant and $\varepsilon=(4-D)/2$.
   At the low-energy scale $\mu=4\pi\,F_0 \approx 1\gev$, the
Wilson coefficients $c_i(\mu)$ get probably corrections
$O(1/N_c,\mu)$ which cannot be calculated exactly until now.
   Therefore, in our approach the coefficients $c_i$, resp. $\xi_i$,
have been treated as phenomenological parameters determined by
experiment from the simultaneous analysis of $K\to2\pi$, $K\to3\pi$
decays \cite{pl2,our}.

   In order to separate the contributions belonging to the
dominating combination $(-\xi_1+\xi_2+\xi_3)$ and to $\xi_4$,
$\xi_5$ respectively, we used again, as in \cite{pl2,our}, the
experimental data on parameters of $K\to2\pi$, $K\to3\pi$ decays given
in table 2: partial decay widths $B_i$ and the expansion coefficients
$g_i$, $h_i$ of the matrix element squared with respect to the
variable $Y$ (see (\ref{expanxy})).
   We obtained the following values of the parameters:
\begin{equation}
(-\xi_1+\xi_2+\xi_3)=  6.96  \pm 0.48,  \qquad
              \xi_4 =  0.516 \pm 0.025, \qquad
              \xi_5 = -0.183 \pm 0.022.
\label{xi}
\end{equation}
   Here we give the errors rescaled with $\chi^2$ according to the
standard procedure \cite{pdg}. The values of $B_i$, $g_i$ and $h_i$,
corresponding to the parameter estimates given below are also
presented in table\,2.
   If we would consider $C_{SP}$ as a free parameter, the
fit fixes its value as $C_{SP}= 1.36 \pm 0.54$, in good agreement
with the prescription (\ref{spprescr}).

   The coefficients $c_i$ written in (\ref{weak-lagr}) cannot be
considered as well defined until a procedure for the bosonization of
hadronic currents is given.
   The procedure used here differs from the original more heuristic one
used in \cite{pl2,our} with respect to some normalization factors and
Fierz-corrections.
   Therefore the numerical values found here cannot be compared
directly to those in \cite{pl2,our}, in spite of the fact that
physical results should nevertheless not disagree, if in both case the
$c_i$ are fixed by the same experimental data.

   As the analysis of the coefficients $c_i$ in leading-log
approximation of QCD has shown, the main contribution to direct
CP-violation comes from the penguin diagrams.
   If we still neglect the contribution of electromagnetic
penguin operators, the imaginary part of the coefficient
$c_5$, responsible for the direct CP-violation, can be
calculated from the relation (\ref{eps}) as
$$ |\im c_5|^{\su{exp}} = 0.053_{-0.011}^{+0.015}\, \epsp .
$$
   This leads to the following estimates for the charge asymmetries of the
slope
parameters
\begin{equation}
|\Delta g(K^\pm \to \pi^\pm \pi^\pm \pi^\mp)| = 0.23_{-0.08}^{+0.05}\,
\epsp \, ,
\qquad
|\Delta g(K^\pm \to \pi^0 \pi^0 \pi^\pm)| = 0.19_{-0.08}^{+0.03}\, \epsp .
\label{delg-values}
\end{equation}
   The difference of the first of these values with respect to the one
given earlier \cite{pl2,our} (less than a factor 2) represents the
model-dependence of this approach.
   The second value, $\Delta g(K^\pm \to \pi^0 \pi^0 \pi^\pm)$, is
further diminished by the effect of \mix{}, formerly not taken into
account in this channel.
   For a final judgement on the predicted $\Delta g$-values, one should
also, however, take into account the effect of electromagnetic
penguins (see below).

   For a comparison of our results with those of \cite{cheng}, we have
to put that estimate into a form comparable to (\ref{delg-values}),
resulting in $\delg \approx 0.03 \, \epsp$.
   The very small numerical value of $\delg$ given in \cite{cheng}
is partly due to the small estimate taken for $\epsp$, based on
calculations of Wilson coefficients using the leading-log
approximation of QCD.
   This estimate is one order of magnitude smaller than the
experimental value used in our papers.
   Clearly, this means that we should indeed explain only the
difference of just one order of magnitude.
   The enhancement of the charge asymmetry of the slope parameter
$\delg$ of about one order of magnitude, compared with the soft-pion
limit estimation, is caused by the contribution connected with
$\Delta^{(1/2, \, 1/2)}$ originating from the different behavior of penguin and
non-penguin isotopic amplitudes $\A^{(i)}_{11}$ and $\B^{(i)}_{11}$.
   This difference arises due to fourth-order corrections, \mix{} and
one-loop corrections which are included in table 1.
   The contribution from $\Delta^{(1/2, \, 1/2)}$ increases the value of $\im
F_1$ by a factor of $19$ (see table $1d$) in comparison to the value
of the soft-pion limit (see table 1a).
   On the other hand, one-loop corrections significantly
decrease the value of $\im F_2$ due to suppression of $\B_{23}$.
   But as a result of both contributions the numerator of the formula
(\ref{delg}) significantly increases in comparison with its
denominator (see values of $\re F_{1,2}$ in table 1).
   So, the additional enhancement of $\delg$ can be traced back to an
interplay of higher-order derivative terms, meson rescattering and \mix.
   In particular, the influence of $p^4$-contributions proved to be decisive.
   It seems impossible to us to analyze this question without detailed
calculations including all the above mentioned corrections (the
$p^4$-corrections have been considered only approximately by
\cite{cheng}, but the one-loop corrections for $K\to3\pi$ decays have
not been calculated in ref.{} \cite{cheng,dambrosio,isidori,shabalin} at all).

   It is worth noting that the corrections to the soft-pion amplitudes of
$K\to2\pi$, $K\to3\pi$ decays, discussed here, essentially modify also
the well-known Li-Wolfenstein relation \cite{li-wolf} which
connects the direct $CP$-violation parameters of
$K_S^0\to\pi^+\pi^-\pi^0$ and $K_L^0\to2\pi$ decays:
$$
\varepsilon '_{+-0} = -2\varepsilon ' .
$$
   As it was discussed in papers \cite{dhv,fg,our} the $p^4$-contributions,
\mix{} and one-loop corrections can essentially enhance the direct
CP-violation effects in the decays $K^0(\bar{K}^0)\to\pi^+\pi^-\pi^0$
as compared with $K^0\to2\pi$ decays and the ratio
$|\varepsilon '_{+-0} / \varepsilon '|$ might be larger than the
Li-Wolfenstein prediction, which corresponds to the soft-pion limit
of chiral theory.
   In our approach \cite{our} we found
$|\varepsilon '_{+-0}| \approx 6.8|\varepsilon '|$.
   Also here, as in the case of charged $K$ decays, the different
behavior of mesonic matrix elements for penguin and nonpenguin
operators at the level of $p^4$-corrections plays a most important role.
   Thus, in the estimation of observable effects of direct CP-violation,
their contributions have therefore to be separated explicitly in any
order.\footnote{The description of $\eta '\to\eta 2\pi$ decays
represents the most striking example of the importance of higher-order
corrections (see \cite{strong} ). In the soft-pion limit the
amplitudes of  $\eta '\to\eta 2\pi$ decays are different from zero
only due to the contribution of chiral symmetry breaking. The
corresponding total width $\Gamma_{\eta '\to\eta 2\pi}^{(soft-pion)}$ =
4 $ keV$ turns out to be much smaller then the experimental value
$\Gamma_{\eta '\to\eta 2\pi}^{(exp.)} = (189 \pm 32) keV$. However, the
$p^4$-corrections strongly increase the total width up to the value
$\Gamma_{\eta '\to\eta 2\pi}^{(p^4)}$ = 220$ keV$ which is in a good
agreement with the experimental data.}

   Finally we tried to estimate the effect of electroweak penguin
operators ${\cal O}_{7,8}$ on the charge asymmetry $\delg$.
   The contributions of ${\cal O}_7$ may be neglected in comparison to the
dominating contributions of the operator ${\cal O}_8$ (see table 1).
   Furthermore, calculations of Wilson coefficients $c_{5,8}$ in
leading-log approximation of QCD \cite{buchalla,paschos} show, that
$\re c_8 \ll \re c_5$, so the contributions of electroweak penguins to the
absolute
value of amplitudes for nonleptonic $K$-decays may be neglected as
well.
   The most important effect of the electroweak penguin operator
${\cal O}_8$ appears in the parameters of direct CP-violation being
discussed here.
   Because of the strong dependence of $\im c_8$ on the
mass of the $t$-quark for $m_t \geq 100$ GeV on one hand, and the weak
dependence of $\im c_5$ on $m_t$ on the other hand, the contribution to
direct CP-violation from electroweak penguin operators becomes
important for large $m_t$.
   Using the dependence of the ratio
$\eta (m_t)=\im c_8 / \im c_5$ on $m_t$, as derived in the papers
\cite{buchalla,paschos},
we repeated the calculational procedure described above.
   As a result we found the phenomenological connection between
$\delg$ and $\varepsilon^{'}$ shown in fig.1, which demonstrates
that taking into account the electroweak penguin operator ${\cal O}_8$
does not only not suppress the effect of direct CP-violation in
$K\to3\pi$ decays, but may lead to an additional enhancement in
comparison with that seen in $K\to2\pi$ decays.
   Of cource, this makes the experimental investigation of
$K\to3\pi$ decays with high statistics even more interesting.

   Two of the authors (A.A.Belkov and A.V.Lanyov) are grateful for the
hospitality extended to them at the DESY-Institute for High Energy
Physics, Zeuthen.
\def\largelinestretch{\renewcommand{\baselinestretch}{0.9}}
\largelinestretch\small\normalsize
\textheight = 1.27 \textheight
\newpage
\begin{center}
                    \Large Figure Caption \large
\end{center}

     Fig.1. Dependence of the asymmetry of the slope parameter $\Delta
g$ on the top quark mass.
\newpage
   %
   %
\begin{center}
                    \Large Table 1. \large
           Isotopic amplitudes of $K\to2\pi$, $K\to3\pi$ decays \\
         under successive inclusion of various corrections \\
\end{center}
   %
   %
{\large                 1a)  Soft-pion limit} \\
   %
\begin{tabular}{|l|*{8}{r}|} \hline\hline
& $\O_1$ & $\O_2$ & $\O_3$ & $\O_4$ & $\O_5$ & $\O_6$ & $\alpha \O_7$ & $\alpha
\O_8$ \\ \hline
$\re A_0$     & -1.000 &  1.000 &  1.000 &  0.000 &-22.42  &  0.000 &  0.015 &
2.362 \\
$\re \A_{11}$ & -1.000 &  1.000 &  1.000 &  0.000 &-22.42  &  0.000 &  0.015 &
-9.343 \\
$\re \B_{11}$ & -1.000 &  1.000 &  1.000 &  0.000 &-22.42  &  0.000 & -0.073 &
-0.903 \\ \hline
$\re A_2$     &  0.000 &  0.000 &  0.000 &  1.000 &  0.00  &  0.000 & -0.015 &
1.017 \\
$\re \A_{13}$ &  0.000 &  0.000 &  0.000 &  1.000 &  0.00  &  0.000 & -0.015 &
-4.471 \\
$\re \B_{13}$ &  0.000 &  0.000 &  0.000 & -1.068 &  0.00  &  0.000 & -0.028 &
-2.891 \\
$\re \B_{23}$ &  0.000 &  0.000 &  0.000 &  6.932 &  0.00  &  0.000 &  0.030 &
-2.289 \\ \hline\hline
\end{tabular}
{}~\begin{tabular}{l}
$\im F_1 =   1.19 \epspp$\\[0.5em]
$\im F_2 = -3.97 \epspp$\\[0.5em]
$\re F_1 = 114.5$\\[0.5em]
$\re F_2 = 40.2$
\end{tabular}
\\[1em]
   %
   %
{\large                 1b) Additional inclusion of $p^4$-corrections} \\
   %
\begin{tabular}{|l|*{8}{r}|} \hline\hline
& $\O_1$ & $\O_2$ & $\O_3$ & $\O_4$ & $\O_5$ & $\O_6$ & $\alpha \O_7$ & $\alpha
\O_8$ \\ \hline
$\re A_0$     & -1.000 &  1.000 &  1.000 &  0.000 &-22.42  &  0.000 &  0.015 &
2.362 \\
$\re \A_{11}$ & -1.217 &  1.217 &  1.217 &  0.000 &-19.22  &  0.000 &  0.018 &
-7.233 \\
$\re \B_{11}$ & -0.701 &  0.701 &  0.701 &  0.000 &-30.45  &  0.000 & -0.077 &
-0.623 \\ \hline
$\re A_2$     &  0.000 &  0.000 &  0.000 &  1.000 &  0.00  &  0.000 & -0.015 &
1.017 \\
$\re \A_{13}$ &  0.000 &  0.000 &  0.000 &  1.217 &  0.00  &  0.000 & -0.018 &
-3.757 \\
$\re \B_{13}$ &  0.000 &  0.000 &  0.000 & -0.914 &  0.00  &  0.000 & -0.030 &
-4.033 \\
$\re \B_{23}$ &  0.000 &  0.000 &  0.000 &  7.386 &  0.00  &  0.000 &  0.024 &
-3.499 \\ \hline\hline
\end{tabular}
{}~\begin{tabular}{l}
$\im F_1 =  10.4 \epspp$\\[0.5em]
$\im F_2 = -3.88 \epspp$\\[0.5em]
$\re F_1 = 125.9$\\[0.5em]
$\re F_2 = 48.1$
\end{tabular}
\\[1em]
   %
   %
{\large                 1c) Additional inclusion of \mix{} } \\
   %
\begin{tabular}{|l|*{8}{r}|} \hline\hline
& $\O_1$ & $\O_2$ & $\O_3$ & $\O_4$ & $\O_5$ & $\O_6$ & $\alpha \O_7$ & $\alpha
\O_8$ \\ \hline
$\re A_0$     & -0.996 &  0.979 &  0.961 & -0.010 &-22.15  &  0.004 &  0.015 &
2.400 \\
$\re \A_{11}$ & -1.243 &  1.100 &  1.012 &  0.080 &-19.82  &  0.036 &  0.017 &
-7.071 \\
$\re \B_{11}$ & -0.788 &  0.641 &  0.323 &  0.051 &-30.61  &  0.037 & -0.076 &
-0.583 \\ \hline
$\re A_2$     & -0.004 &  0.021 &  0.039 &  1.009 & -0.27  & -0.004 & -0.015 &
0.980 \\
$\re \A_{13}$ &  0.056 &  0.109 &  0.467 &  1.287 & -1.22  & -0.041 & -0.021 &
-3.588 \\
$\re \B_{13}$ &  0.069 &  0.123 &  0.512 & -0.879 & -0.52  & -0.048 & -0.033 &
-4.035 \\
$\re \B_{23}$ & -0.019 &  0.063 &  0.134 &  7.471 & -0.67  & -0.011 &  0.022 &
-3.462 \\ \hline\hline
\end{tabular}
{}~\begin{tabular}{l}
$\im F_1 =  9.8 \epspp$\\[0.5em]
$\im F_2 = -4.09 \epspp$\\[0.5em]
$\re F_1 = 131.7$\\[0.5em]
$\re F_2 = 52.6$
\end{tabular}
\\[1em]
   %
   %
{\large                 1d) Additional inclusion of one-loop corrections} \\
   %
\begin{tabular}{|l|*{8}{r}|} \hline\hline
& $\O_1$ & $\O_2$ & $\O_3$ & $\O_4$ & $\O_5$ & $\O_6$ & $\alpha \O_7$ & $\alpha
\O_8$ \\ \hline
$\re A_0$     & -1.258 &  1.241 &  1.205 & -0.010 &-27.68  &  0.004 &  0.034 &
3.622 \\
$\re \A_{11}$ & -1.273 &  1.129 &  1.187 &  0.080 &-17.47  &  0.076 & -0.009 &
-8.084 \\
$\re \B_{11}$ & -0.441 &  0.293 &  0.565 &  0.051 &-48.38  &  0.037 &  0.003 &
-8.091 \\ \hline
$\im A_0$     & -0.461 &  0.461 &  0.461 &  0.000 &  9.78  &  0.000 &  0.007 &
-0.506 \\
$\im \A_{11}$ & -0.087 &  0.087 &  0.087 &  0.000 & -4.29  &  0.000 &  0.003 &
-1.167 \\
$\im \B_{11}$ & -1.104 &  1.104 &  1.104 &  0.000 &-48.45  &  0.000 &  0.009 &
-7.063 \\ \hline
$\re A_2$     & -0.004 &  0.021 &  0.039 &  0.692 & -0.27  & -0.004 & -0.013 &
0.485 \\
$\re \A_{13}$ &  0.056 &  0.109 &  0.467 &  1.798 & -1.22  & -0.081 & -0.041 &
-5.592 \\
$\re \B_{13}$ &  0.069 &  0.123 &  0.506 & -0.710 & -0.52  & -0.048 & -0.038 &
-4.967 \\
$\re \B_{23}$ & -0.019 &  0.063 &  0.134 &  1.625 & -0.68  & -0.011 & -0.007 &
-0.515 \\ \hline
$\im A_2$     &  0.000 &  0.000 &  0.000 & -0.231 &  0.00  &  0.000 &  0.003 &
-0.231 \\
$\im \A_{13}$ &  0.000 &  0.000 &  0.000 &  0.184 &  0.00  &  0.000 & -0.003 &
-0.593 \\
$\im \B_{13}$ &  0.000 &  0.000 &  0.000 &  0.956 &  0.00  &  0.000 & -0.017 &
-3.444 \\
$\im \B_{23}$ &  0.000 &  0.000 &  0.000 & -1.018 &  0.00  &  0.000 & -0.005 &
0.272 \\ \hline\hline
\end{tabular}
{}~\begin{tabular}{l}
$\im F_1 =  22.6 \epspp$\\[0.5em]
$\im F_2 = -0.63 \epspp$\\[0.5em]
$\re F_1 = 143.9$\\[0.5em]
$\re F_2 = 14.0$
\end{tabular}
   %
   %
\newpage
   %
   %
\begin{center}
                    \Large Table 2. \large
           Experimental and theoretical values of the parameters \\
              for $K\to2\pi$, $K\to3\pi$ decays
\end{center}
   %
   %
\begin{center}
\begin{tabular}{|l|@{\hspace{7mm}}r@{$\pm$}l@{\hspace{10mm}}c@{\hspace{5mm}}|}
 \hline\hline \rule[-2.5ex]{0ex}{5ex}
 Parameter & \multicolumn{2}{l}{Experiment [16]}&Theory\\
\hline\rule{0ex}{2.6ex}%
{}~~~$B_{+0}$    &   0.2117 & 0.0015   & 0.2125 \\
{}~~~$B_{+-}$    &   0.6861 & 0.0024   & 0.6965 \\
{}~~~$B_{00}$    &   0.3139 & 0.0024   & 0.3130 \\[0.5ex]
\hline\rule{0ex}{2.6ex}%
{}~~~$B_{++-}$   &   0.0559 & 0.0005   & 0.0553 \\
{}~~~$g_{++-}$   &  -0.2162 & 0.0031   &-0.2144 \\
{}~~~$h_{++-}$   &   0.011  & 0.005    & 0.016  \\[0.5ex]
\hline\rule{0ex}{2.6ex}%
{}~~~$B_{00+}$   &   0.0173 & 0.0004   & 0.0176 \\
{}~~~$g_{00+}$   &   0.594  & 0.019    & 0.552  \\
{}~~~$h_{00+}$   &   0.035  & 0.015    & 0.014  \\[0.5ex]
\hline\rule{0ex}{2.6ex}%
{}~~~$B_{+-0}$   &   0.1238 & 0.021    & 0.1182 \\
{}~~~$g_{+-0}$   &   0.670  & 0.014    & 0.648  \\
{}~~~$h_{+-0}$   &   0.079  & 0.007    & 0.151  \\[0.5ex] \hline\hline
\end{tabular}
\end{center}
   %
   %
\end{document}